\documentclass[12pt,a4paper]{article}
\usepackage[latin1]{inputenc}
\usepackage{amsmath,graphicx,cite}
\usepackage{amsfonts}
\usepackage{amssymb}

\begin{document}

\begin{center}
{\Large\bf Implications of the Crystal Barrel data for meson-baryon symmetries}\\
\bigskip
{\bf S. S. Afonin\footnote{On leave of absence from V. A. Fock Department of
Theoretical Physics, St. Petersburg State
University, 1 ul. Ulyanovskaya, 198504 St. Petersburg, Russia.}}\\
\smallskip
University of Bochum, Department of Physics and Astronomy,
Theoretical Physics II, 150 Universit\"{a}tsstrasse, 44780 Bochum,
Germany
\end{center}

\begin{abstract}
Making use of numerous resonances discovered by the Crystal Barrel
Collaboration we discuss some possible relations between the
baryon and meson spectra of resonances composed of the light
non-strange quarks. Our goal is to indicate new features that
should be reproduced by the realistic dynamical models describing
the hadron spectrum in the sector of light quarks.
\end{abstract}


The observation of more than thirty new inelastic $\bar{p}p$
resonances in the recent analyses of Crystal Barrel and PS172
data~\cite{ani,ani2,bugg,klempt}
has been a spectacular event in the spectroscopy of light
hadrons below 2.4 GeV. It is well known that the hadron mass
spectrum contains an important information on properties of the strong
interactions. In this regard the data obtained remarkably confirmed
various spectral regularities that indicate to a high degree of
symmetry emerging in the large distance strong
interactions~\cite{kaid,epj,ej,ej_b,ej_c,sh,ej2}. And what is more the
concept of Regge trajectories acquired a more solid experimental
support. It is tempting to assume that instead of using the
language of hadron resonances one could discuss some important
phenomena in QCD, say the chiral symmetry breaking, in the
language of geometrical behavior of Regge trajectories. We will
try to demonstrate how the latter language may be used.

The purpose of this note is to bring attention to possible
implications of the Crystal Barrel data for meson-baryon
symmetries. We expect that the joint consideration of meson and
baryon sectors could be useful for both sectors of hadron
spectroscopy.

\begin{figure*}
\vspace{-0 cm}
\hspace{-0 cm}
\resizebox{1\textwidth}{!}{%
  \includegraphics{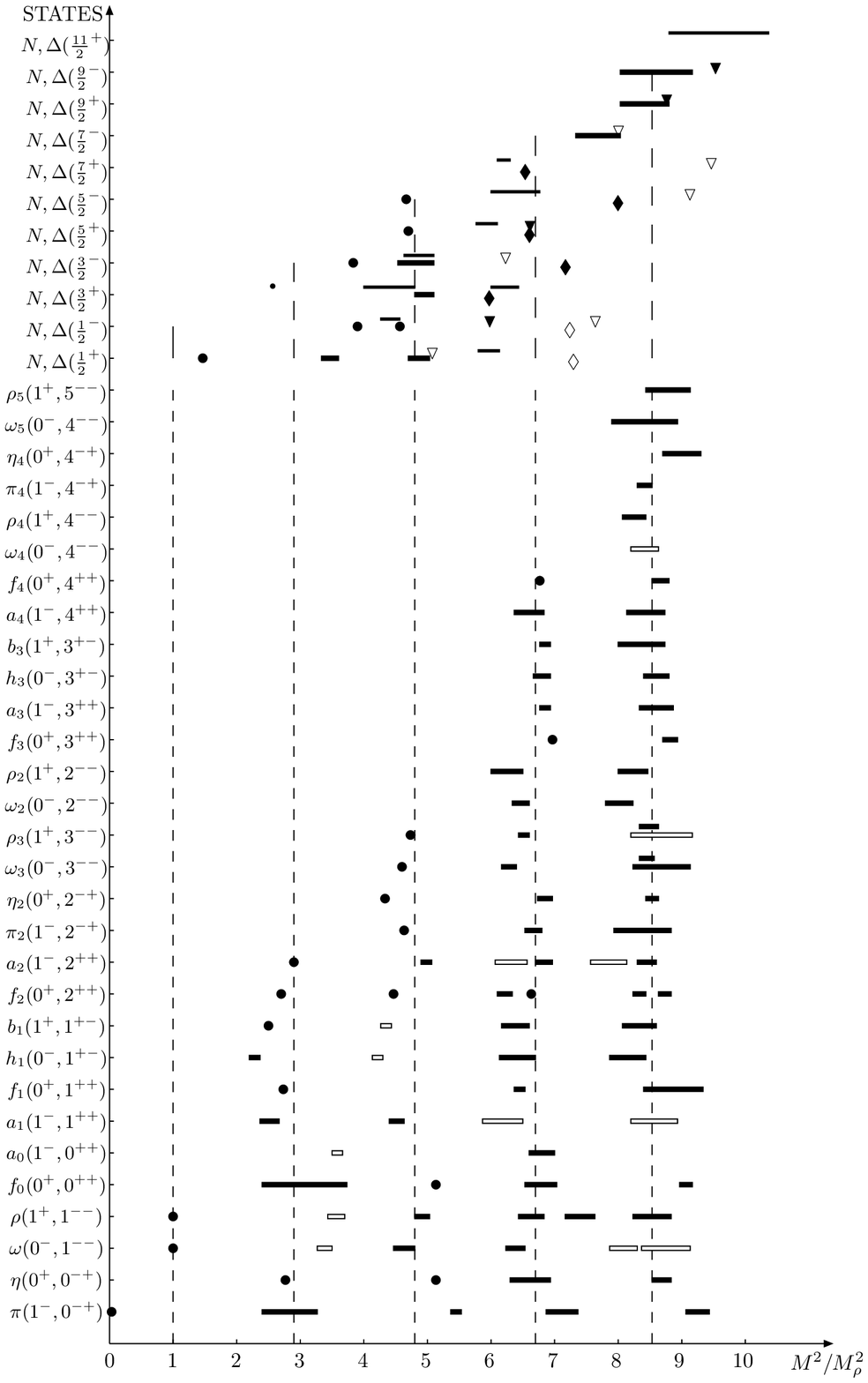}
}
\vspace{-4 cm}
\caption{The spectrum of light non-strange hadrons in units of $M_{\rho(770)}^2$.
The data for mesons is taken from refs.~\cite{pdg} and~\cite{bugg}
(for the last two clusters) and for baryons from ref.~\cite{pdg}.
Experimental errors are indicated except the one- and two-star baryons
(by thin strips for $\Delta$).
Circles stay when errors are negligible. For mesons the dashed lines mark the
mean (mass)$^2$ in each cluster of states (these lines are continued to the baryon
sector) and the open strips denote the one-star states from ref.~\cite{bugg}
or the states that are dubious as non-strange mesons.
The symbols $\blacklozenge$ and $\lozenge$ denote the two- and one-star nucleons
correspondingly. The symbols $\blacktriangledown$ and $\triangledown$ do the same
for $\Delta$-baryons.}
\label{f0}
\end{figure*}

As a starting point for our discussions we present Fig.~\ref{f0}
where the combined meson-baryon light non-strange spectrum is
displayed on one plot. For the details of meson plot we refer
to~\cite{epj}, that plot is supplemented in Fig.~\ref{f0} with the
baryon states.

A prominent feature of light non-strange meson spectrum is the
well-pronounced clustering of states near equidistant values of
masses square~\cite{epj,ej,ej_b,ej_c}. The corresponding baryon sector is
known to reveal a tendency to clustering as well. Our first
observation is that the position of the third meson cluster
happens to coincide with the first more or less pronounced cluster
of baryons. Most of states in both clusters are well
established~\cite{pdg}, thus this is a quite secure effect. The
resonances in the second baryon cluster (especially the well
established ones) are shifted towards lower energies with respect
to the fourth meson cluster. It seems that there is no enough data
to speak about the next baryon cluster that would correspond to
the fifth meson one.

These observations have important consequences for a possible
dynamical supersymmetry that was suggested long
ago~\cite{anselmino}. Such a meson-baryon symmetry is commonly
motivated by the quark-diquark structure of baryons where the
diquark might have approximately the same constituent mass as
quark and behave like antiquark in mesons. In particular, in
ref.~\cite{wil,wil2} it is suggested that the diquarks can become well defined
objects in the highly excited baryons due to large separation from
the quark inside a baryon. The masses of some mesons and baryons
are indeed surprisingly close. Unfortunately, such coincidences
prove nothing because one should analyze the full amount of data.
As seen in Fig.~\ref{f0} near 1.7~GeV (the third meson cluster)
some kind of meson-baryon supersymmetry can indeed take place. But
the resonances obtained by the Crystal Barrel Collaboration (they
mainly constitute the last two clusters in Fig.~\ref{f0}) do not
confirm convincingly this symmetry. Even if one assumes that there
are different kinds of diquarks (the spin singlet and spin triplet
ones) with different masses and the constituent mass of diquarks
depend on the energy scale, any meson-baryon symmetry should
dictate a certain correlation between the number of mesons and
baryons and this correlation should recur at higher energies. One
indeed observes some recurrences in both meson and baryon sectors
--- the cluster structure of spectrum --- but looking at the third
and fourth meson clusters and their baryonic counterparts it is hard
to see any clear-cut correlation in the number of states. It is
not excluded of course that the reason is just a lack of data in
the baryon sector.

In what follows we will indicate some new features of meson-baryon
spectra that likely should be reproduced in any viable dynamical models
describing the spectrum of light non-strange hadrons.

The recurrence patterns in meson spectrum seems to suggest that
the MacDowell symmetry is (partly) realized in the meson sector.
We will remind briefly the essence of this symmetry.
Consider a Regge trajectory $\alpha(s)$. According to Regge-pole
theory~\cite{novozhilov}, in general there will be a series of daughter trajectories
$\alpha_k(s)$ in the angular momentum plane, of alternating
signature, satisfying
\begin{equation}
\alpha_k(s)=\alpha_0(s)-k,\qquad k=1,2,\dots.
\end{equation}
The MacDowell symmetry says that all baryon trajectories with
equal isospin, the same signature, and opposite parity are
degenerate. This property was indeed observed in the late 1960s~\cite{bar,bar2}.

The meson spectrum also reveals the parity doubling and the
Crystal Barrel data remarkably confirmed this phenomenon
(see~\cite{ej2} for a review).
A feature of meson trajectories is that the exchange degeneracy,
resulting from the absence of $I=2$ mesons~\cite{novozhilov},
leads to the approximate coincidence of trajectories formed by
$\rho$, $\omega$, $f_2$, and $a_2$ resonances. As a result, the
$\rho$ ($\omega$) trajectory with $J=1,3,5,\dots$
and $a_2$ ($f_2$) trajectory of opposite signature with $J=2,4,6,\dots$
coalesce into one master trajectory (we recall that a reggeized pole amplitude
of negative signature has poles at odd $J$ and that of positive signature
does at even $J$). The equidistant sequence of
daughters gives rise to linear "radial" trajectories --- the
towers of states with the same quantum numbers, the "radial"
excitations in the language of potential models. For instance, the
$J=1$ states on the daughters of master trajectory have the
quantum numbers of the $\rho$ ($\omega$) meson. In addition, the
Adler self-consistency condition in dual models
requires the master trajectory to have the intercept
$\alpha_0(m_{\pi}^2)=\frac12$ and, hence, the slope
$\alpha'=(2m_{\rho}^2-2m_{\pi}^2)^{-1}$~\cite{novozhilov}.

\begin{figure*}
\vspace{-4 cm}
\hspace{-2 cm}
\resizebox{1.2\textwidth}{!}{%
  \includegraphics{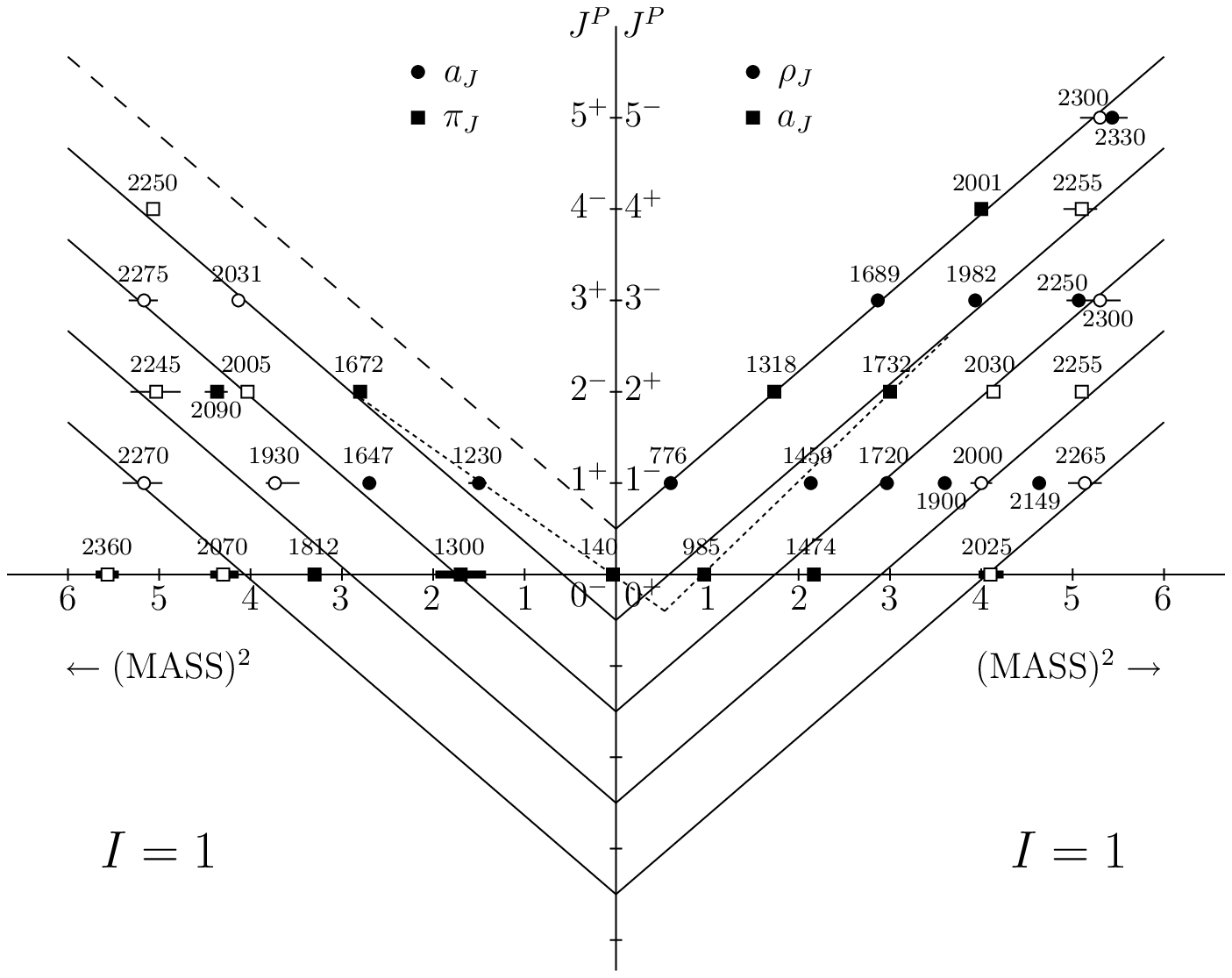}
}
\vspace{-10 cm}
\caption{The trajectories of isovector non-strange mesons (in GeV$^2$).
The filled circles (squares) denote the states
contained in the PDG~\cite{pdg}. The open circles (squares) are the
resonances observed in the Crystal Barrel experiment~\cite{bugg}
(they are usually cited by the PDG in section "Further States").
The averaged values of masses are indicated in MeV and the experimental errors
(if significant) are shown. The dashed line is the absent
MacDowell pair for the leading master trajectory. The dotted line imitates the
deviations presumably caused by CSB.}
\label{f1}
\end{figure*}

\begin{figure*}
\vspace{-4 cm}
\hspace{-2 cm}
\resizebox{1.2\textwidth}{!}{%
  \includegraphics{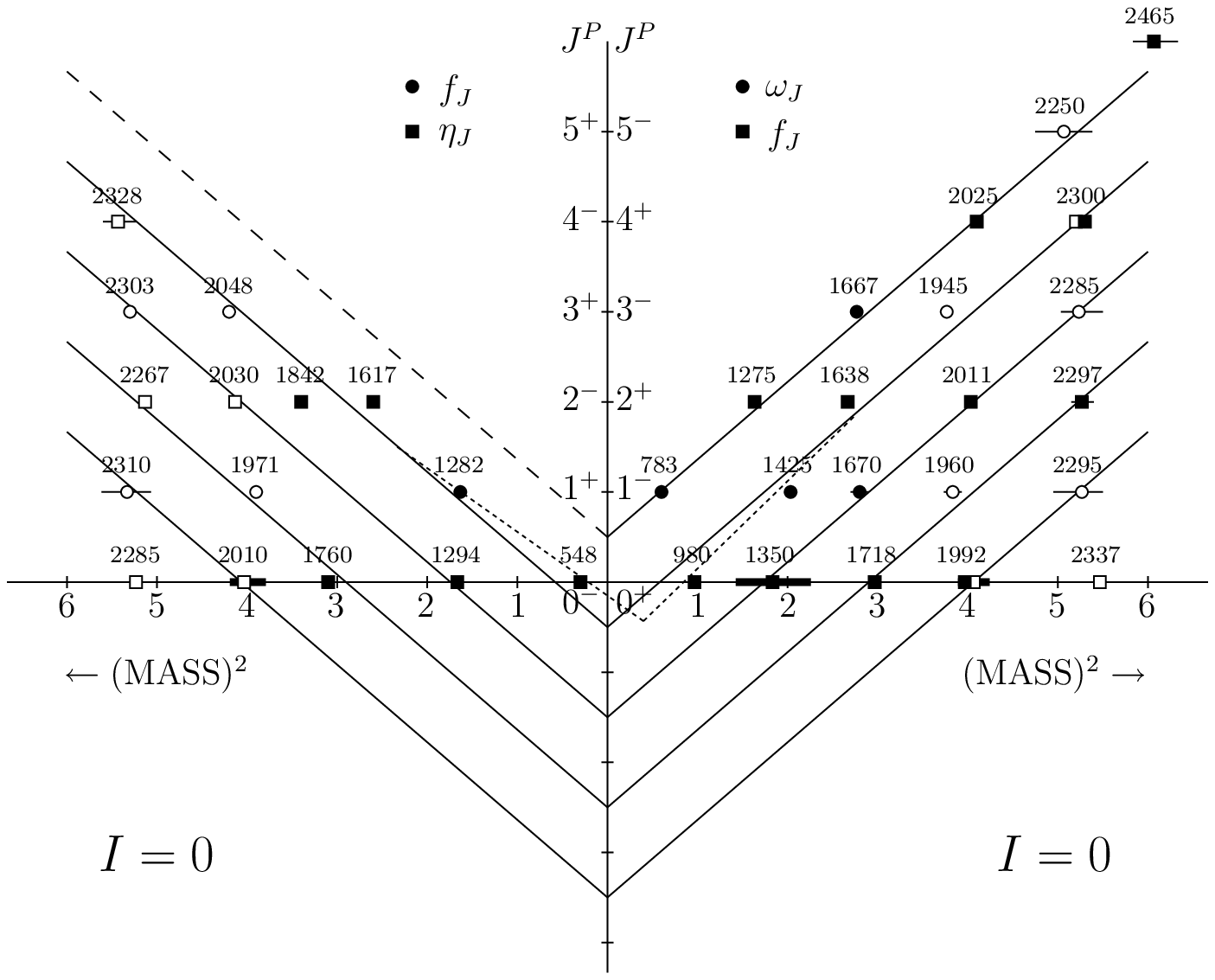}
}
\vspace{-10 cm}
\caption{The trajectories of isoscalar non-strange mesons.
The notations are as in Fig.~\ref{f1}.}
\label{f2}
\end{figure*}

The results of Regge-pole theory are general consequences of
unitarity and analyticity. The role of QCD is to provide the
linearity of trajectories and the scale $\alpha'$. The
corresponding trajectories are displayed in Fig.~\ref{f1} and
Fig.~\ref{f2}. The MacDowell trajectory pairs have to be joined
at the point $M^2=0$, for convenience in representation the
remaining trajectories are joined in a linear manner (like the
baryon trajectories in ref.~\cite{bar,bar2}).
We assign the experimental light unflavored
states to the drawn trajectories according to their averaged masses and
quantum numbers. The Crystal Barrel experiment revealed many new
resonances in the energy interval 1.9 - 2.4 GeV,
they occupy the vacant places in Fig.~\ref{f1} and Fig.~\ref{f2}
with surprisingly high accuracy. In addition, in all cases when
they do not coincide with the known states from the Particle Data Group (PDG)~\cite{pdg},
the agreement with the theoretical expectations is
improved. As a result one observes that
\emph{all \underline{meson} trajectories are approximately MacDowell symmetric
except the leading master trajectory}.

The first daughter of master trajectory and its MacDowell
symmetric trajectory, the leading pseudoscalar one, show up an
intriguing pattern of deviations from exact linearity for the low
lying states (see Fig.~\ref{f1} and Fig.~\ref{f2}). Presumably the
deviations are caused by the chiral symmetry breaking (CSB) at low
energies. It can be shown that the masses square of shifted
states behave approximately as if the total length of joined
trajectory remained unchanged after deviations, {\it i.e.} the
joined trajectories are approximately "rigid" with respect to the
strong interaction dynamics. In Fig.~\ref{f1}
and Fig.~\ref{f2} we have confined ourselves by the linear form of
deviations, but the conclusion holds beyond this simplification.
The observed correlations imply, in particular, that there may be
a hidden relation between CSB and the violation of OZI-rule for scalar
mesons ({\it i.e.} a considerable mixture of strange and non-strange
components).

\begin{figure*}
\vspace{-4cm}
\hspace{-1cm}
\resizebox{1.1\textwidth}{!}{%
  \includegraphics{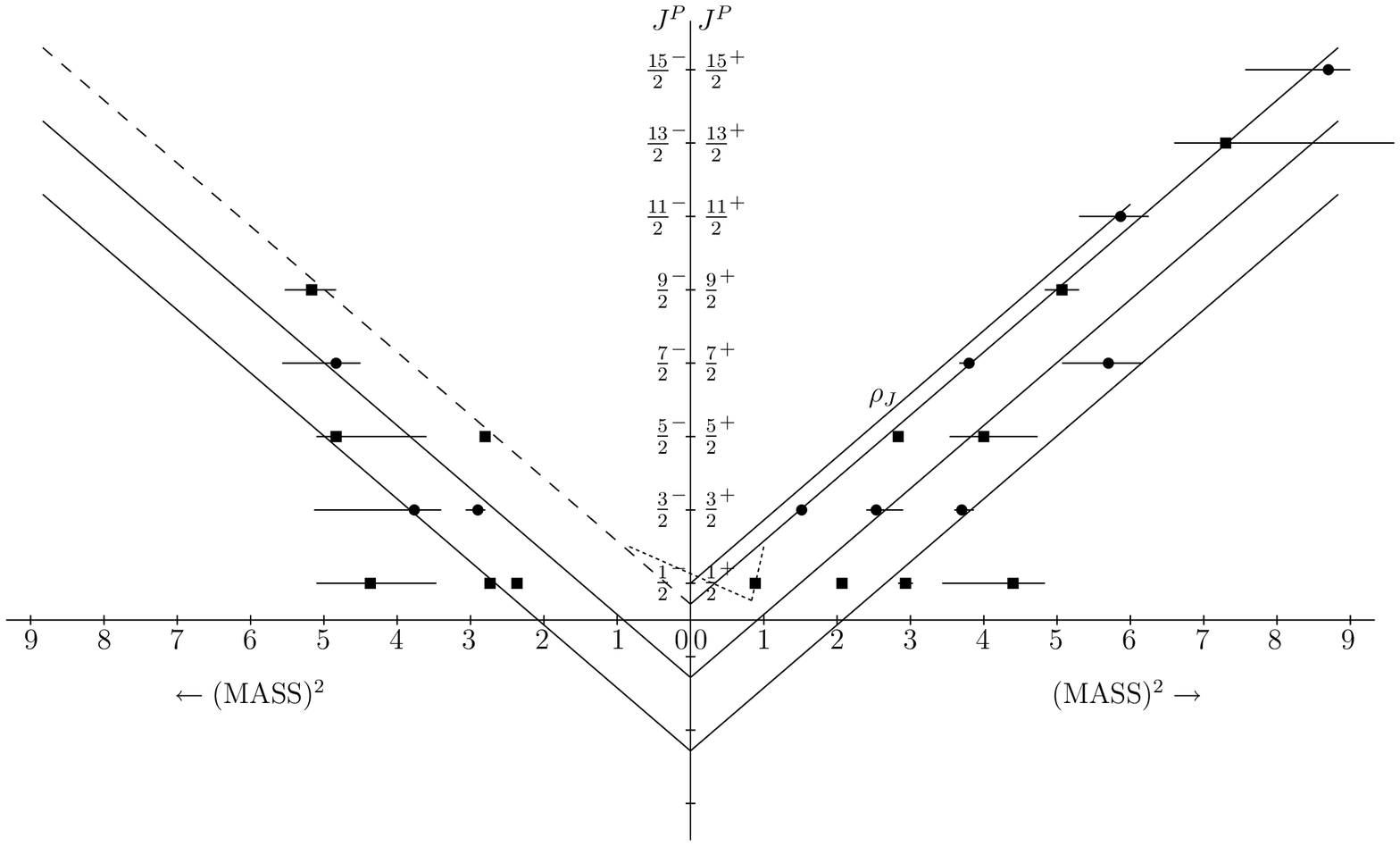}
}
\vspace{-8cm}
\caption{The Regge trajectories of $N_{\frac12^+}$ (squares)
and $\Delta_{\frac32^+}$ (circles) and their daughters. The experimental
errors are indicated (for some states the Particle Data does not
provide errors, in these cases we plot the whole range of masses
reported for a given state, the ensuing errors are typically
quite large). The leading $\rho$-meson trajectory from Fig.~\ref{f1}
is drawn for comparison. The dashed line is the MacDowell symmetric pair
for the leading baryon trajectory. The dotted line shows a possible pattern
of deviation from linearity at low energies for the leading nucleon trajectories.}
\label{f3}
\end{figure*}

In Fig.~\ref{f3}, we display the spectrum of leading nucleon and
delta trajectories and their daughters. The slopes of baryon and
meson trajectories are known to coincide, for comparison we present
in Fig.~\ref{f3} the leading $\rho$-meson trajectory from
Fig.~\ref{f1}. In addition, the leading nucleon and delta
trajectories seem to coalesce into one baryon master trajectory.
Similarly to the mesons, the leading delta trajectory does not
possess the MacDowell symmetric pair, but this is not the case
for the leading nucleon trajectory.

If an approximate symmetry between the meson and baryon spectra
takes place it can provide new inside into the phenomenon of CSB.
The deviations of trajectories from linearity at low energies are
believed to be caused mainly by CSB in QCD. What happens to the
spectrum of unflavored hadrons if we could somehow restore the chiral
symmetry, maintaining the other features of confining QCD? One can
naturally expect that the meson master trajectory remains almost
unaffected as long as the mass of $\rho$-meson originates, most
probably, from the QCD mass gap. On the other hand,
in the case of exact meson-baryon
supersymmetry, the leading baryon trajectories should coincide with
the meson master trajectory
(as seen from Fig.~\ref{f3} in reality there is a constant shift
between these trajectories, this fact is interesting by
itself and is not yet understood). Since
the intercept of the meson master trajectory is approximately
$\frac12$, the ground nucleon becomes massless in the chirally symmetric
world. The fact that the mass of ground nucleon is mostly
generated by CSB in QCD, has been known
for long ago~\cite{ioffe,ioffe2}, we just emphasize its relation with a
possible meson-baryon symmetry. In addition, assuming a similar
pattern of deviation from linearity as in the meson sector, one
can immediately see from Fig.~\ref{f3} why the ground nucleon state
does not have a parity partner --- the latter becomes tachyon
after CSB and disappears from the physical spectrum. A dynamical
description of this mechanism is a challenge for models describing
CSB and hadron spectrum.

Another novel indication on the possible existence of meson-baryon
symmetry could be the fact that the light non-strange spectrum of
both mesons and baryons is, in a group-theoretical sense, similar
to the spectrum of the hydrogen atom, but this is a subject of a
separate paper~\cite{prc3}.

Concluding this note, we hope that a new way of representation of
data on light non-strange hadrons proposed here and ensuing new
features of spectra will rise interest to the related problems
both from the theoretical and from the experimental side.

The work was supported by the Ministry of Education of
Russian Federation, grant RNP.2.1.1.1112, by the Government
of Sankt-Petersburg, grant 29-04/23, and by the Alexander von
Humboldt Foundation.

\newpage

\newpage

\newpage

\end{document}